\documentclass[sigconf]{acmart}

\copyrightyear{2026}
\acmYear{2026}
\setcopyright{cc}
\setcctype{by}
\acmConference[EAAMO '26]{Equity and Access in Algorithms, Mechanisms, and Optimization}{November 05--07, 2026}{Munich, Germany}
\acmBooktitle{Equity and Access in Algorithms, Mechanisms, and Optimization (EAAMO '26), November 05--07, 2026, Munich, Germany}
\acmDOI{10.1145/3846167.3848608}
\acmISBN{979-8-4007-3015-3/2026/11}

\AtBeginDocument{\providecommand\BibTeX{{\normalfont B\kern-0.5em{\scshape i\kern-0.25em b}\kern-0.8em\TeX}}}

\graphicspath{{figures/}}

\begin{document}

\title{Who Gets Heeded? An Obligation-Level Audit of Responsiveness in EPA Rulemaking}

\author{Jianing Fan}
\authornote{Both authors contributed equally to this research.}
\affiliation{%
  \institution{Columbia University}
  \city{New York}
  \state{NY}
  \country{USA}}
\email{jf3774@columbia.edu}

\author{Yue Yao}
\authornotemark[1]
\affiliation{%
  \institution{Columbia University}
  \city{New York}
  \state{NY}
  \country{USA}}
\email{yy3462@columbia.edu}

\renewcommand{\shortauthors}{Fan and Yao}

\begin{abstract}
Notice-and-comment rulemaking gives any affected party the same formal right to influence federal regulation, but formal access is not substantive capacity to shape rule text. Existing measurement strategies operate at the rule or aggregate-corpus level, too coarse to capture the discrete regulatory obligations where commenters seek change. We introduce \textbf{obligation-level responsiveness auditing}, an auditable, AI-assisted framework for measuring whether public-comment engagement co-occurs with changes to specific regulatory obligations. The framework extracts proposed and final-rule obligations, matches comments to the obligations they address, and classifies proposed-to-final outcomes; we test each of these AI-assisted components against blind human judgment. We apply the framework to 70,075 comments across 36 U.S. Environmental Protection Agency (EPA) anchor rulemakings, drawn from a corpus of 786,197 comments across 6,145 dockets from 2010--2022. Three descriptive findings emerge. First, engagement is associated with revision across rulemakings, but not clearly within the same rulemaking. Second, support-versus-opposition direction does not clearly differentiate outcomes, an informative null inconsistent with simple preference aggregation. Third, under a permissive reconstruction of commenter type, organizational-majority engagement concentrates in editorial-refinement rather than substantive-modification outcomes across rulemakings. A blind human audit of the outcome labels behind this finding preserves it under corrected labels and reveals that text-similarity methods are insufficient for distinguishing editorial from substantive regulatory change. Together, these findings locate the equity asymmetry upstream of agency response: in which rulemakings and obligations different commenter populations have the capacity to identify, interpret, and contest. Obligation-level auditing makes this upstream inequity visible.
\end{abstract}



\maketitle

\section{Introduction}
\label{sec:intro}

Public participation in federal rulemaking is one of the central democratic mechanisms that hold the U.S. administrative state accountable to the people it regulates. Under the Administrative Procedure Act, an agency proposing a binding regulation must publish a Notice of Proposed Rulemaking, accept written comments from any affected party, and consider those comments before finalizing the rule; courts can then review the final rule for arbitrary or capricious disregard of the record. In principle, an individual community member has the same formal right to influence regulatory obligations as a regulated industry.

Procedural access, however, does not guarantee substantive responsiveness. Three decades of empirical work have asked whether public comments change rule text and whether organized interests have systematic advantages over individual commenters, but the question has resisted clean answers because existing measurement strategies operate at granularities too coarse to capture the unit at which commenters seek change. Rule-document outcomes (whether a final rule issued, whether the preamble cites specific commenters) collapse hundreds of binding obligations into a single binary~\cite{yackee2006bias}. Aggregate sentiment shifts cannot distinguish responsiveness from independent agency reconsideration. Both lenses are also too aggregated to support equity comparisons across commenter types, because individual commenters and organizational interests rarely engage the same obligations in the same rulemakings.

The unit that commenters actually contest in notice-and-comment rulemaking is the \textbf{regulatory obligation}: thresholds, exemptions, deadlines, reporting duties, monitoring requirements, and compliance burdens. A 200-page proposed rule contains tens to hundreds of discrete obligations, each potentially the focus of commenter argument, and the agency can preserve each one verbatim, refine it editorially, modify it substantively, or drop it between the proposed and final rule, independent of what happens to the rest of the document. Without obligation-level measurement, we cannot distinguish formal access to rulemaking from the substantive capacity to identify, interpret, and contest the obligations where regulatory change actually occurs. Measuring responsiveness at this level matches the institutional grain at which commenters seek change and, crucially for the equity question, makes visible whether different commenter populations have systematically different capacity to engage different obligations.

\begin{figure*}[!htbp]
\centering
\includegraphics[width=\textwidth]{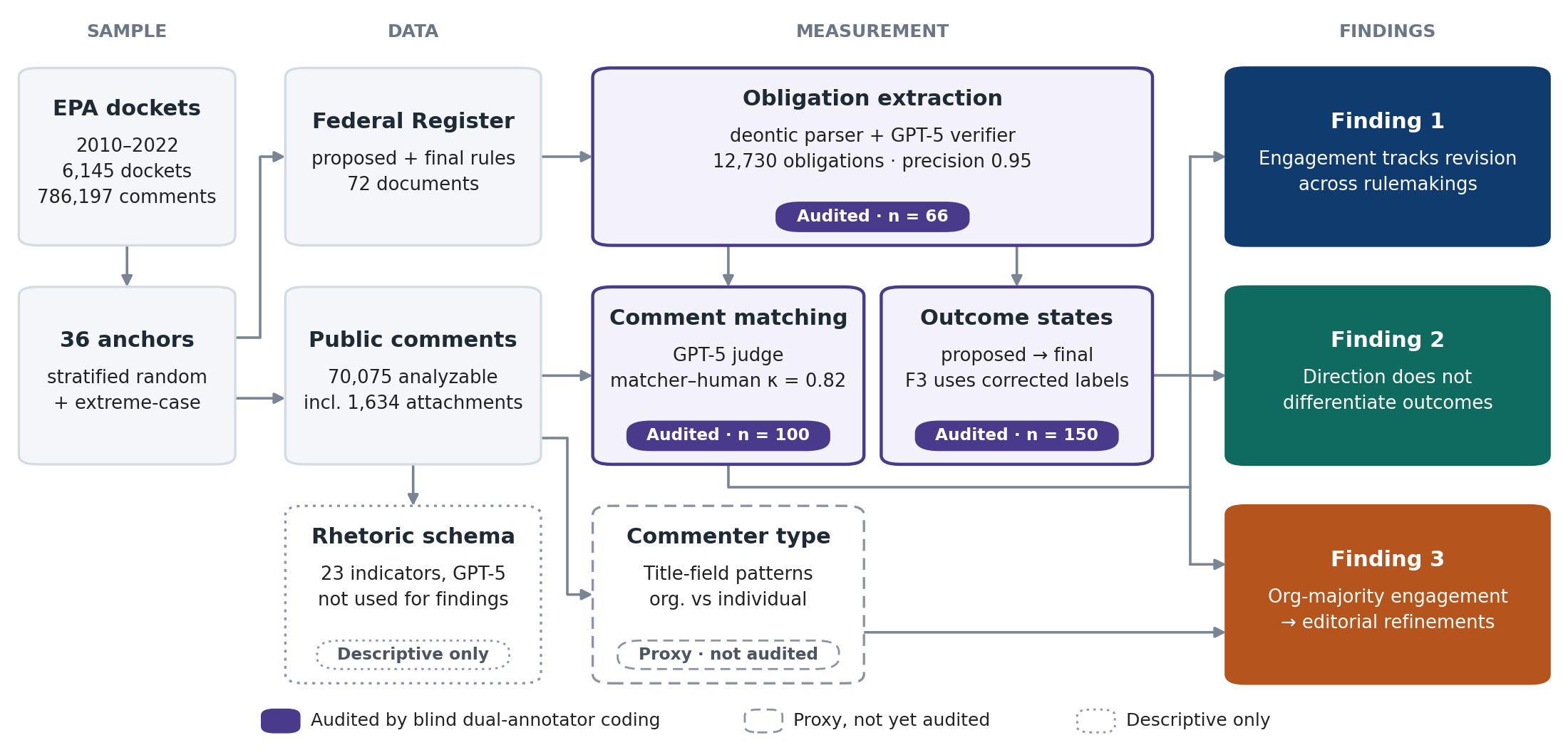}
\caption{The obligation-level responsiveness auditing framework. Each measurement component is marked by its audit status: audited by blind dual-annotator coding (extraction and matching in \S\ref{sec:obligations}, with details in Appendix~\ref{app:validation}; outcome states in \S\ref{sec:results-audit}), a proxy not yet audited (commenter type, \S\ref{sec:methods-commenter}), or descriptive only (the 23-indicator rhetoric schema, Appendix~\ref{app:schema}). Arrows show which components feed each finding; each finding has the color of its result figures.}
\Description{Flow diagram in four columns. Sample: EPA dockets from 2010 to 2022 (6,145 dockets, 786,197 comments) lead to 36 anchor rulemakings chosen by stratified random and extreme-case sampling. Data: Federal Register proposed and final rules (72 documents) and public comments (70,075 analyzable, including 1,634 recovered attachments); below the comments, a 23-indicator rhetoric schema coded with GPT-5 is marked descriptive only and not used for findings. Measurement: obligation extraction (deontic parser plus GPT-5 verifier, 12,730 obligations, precision 0.95, audited with n = 66) feeds comment matching (GPT-5 judge, matcher-human kappa 0.82, audited with n = 100) and outcome states (proposed to final, with Finding 3 using the audit-corrected labels, audited with n = 150); commenter type from Title-field patterns is marked as a proxy that is not yet audited. Findings: comment matching and outcome states feed Findings 1, 2, and 3, and commenter type also feeds Finding 3. Finding 1: engagement tracks revision across rulemakings. Finding 2: direction does not differentiate outcomes. Finding 3: organizational-majority engagement concentrates in editorial refinements. Audited components are outlined in purple, and Findings 1, 2, and 3 are shown in blue, teal, and amber, the colors of their result figures.}
\label{fig:pipeline}
\end{figure*}

This paper introduces \textbf{obligation-level responsiveness auditing} for U.S. notice-and-comment rulemaking. We assemble a population-scale corpus of 786,197 public comments to the U.S. Environmental Protection Agency (EPA) across 6,145 rulemaking dockets from 2010 to 2022, select 36 anchor rulemakings via stratified random and extreme-case sampling, and construct a 70,075-comment analytic sample within those anchors. For each anchor, we extract obligations from both proposed and final Federal Register text via a pipeline that combines a structural deontic parser with a large language model (LLM) verifier (extending Leahey~\cite{leahey2026one}), match public comments to the specific obligations they substantively address, and compare each proposed obligation with its final-rule outcome (Figure~\ref{fig:pipeline}). We evaluate the LLM-assisted extraction and matching components, as well as the text-similarity outcome classifier, through blind human audits; our reconstruction of commenter type remains a proxy, and we report throughout which labels we have audited and which we have not.

Our central claim is that the equity asymmetry in notice-and-comment rulemaking lies \textit{upstream} of agency response. Across rulemakings, engagement tracks revision, and organizational-majority engagement concentrates in editorial refinements; within a single rulemaking, neither the amount of engagement nor the type of commenter is clearly associated with how an obligation changes. The disparity we document is therefore upstream-structural: it operates through which rulemakings and obligations each commenter population has the capacity to engage. Three descriptive findings support this claim, each engaging a default assumption in the regulatory-responsiveness literature.

\paragraph{Finding 1: Engagement is associated with revision across rulemakings, but not clearly within them.} Across 12,243 proposed-side obligations from the 29 anchor rulemakings that yield verified obligations, those addressed by five or more commenters (\textit{high-engagement} obligations) are 17.5 percentage points more likely to be revised between the proposed and final rule than less-addressed obligations (69.0\% versus 51.5\%; $\chi^2 = 19.71$, $p < 0.001$). Within the same rulemaking, however, the association is smaller and statistically inconclusive (odds ratio OR $= 1.54$, 95\% confidence interval CI $[0.83, 2.87]$, $p = 0.174$), so we cannot separate a within-rulemaking effect from differences between rulemakings. Most revisions occur on obligations that no commenter addressed; this pattern is inconsistent with strong-version responsiveness accounts that treat public comments as a primary driver of obligation-level outcomes.

\paragraph{Finding 2: The direction of engagement does not differentiate outcomes.} Obligations addressed by at least one opposing commenter ($n = 212$) are revised at a rate similar to those addressed by at least one supporting commenter ($n = 312$): 62.3\% versus 68.3\% ($p = 0.155$). With these sample sizes the analysis would be unlikely to miss a large directional difference, so we interpret the null as substantively informative: simple preference-aggregation models of agency responsiveness, in which mounting opposition or support should track revision~\cite{yackee2006bias}, are not consistent with it. We address the overlap between the two groups in \S\ref{sec:results-direction}.

\paragraph{Finding 3: Across rulemakings, organizational-majority engagement concentrates in editorial refinements.} Among the 116 high-engagement obligations that were either editorially refined or substantively modified, organizational-majority commenter composition co-occurs with editorial refinement at a substantially higher rate than with substantive modification (87\% versus 57\% organizational-majority under audit-corrected labels; \S\ref{sec:results-equity}). A blind dual-annotator audit of these obligations preserves the direction of the disparity under corrected labels while reducing statistical precision, and reveals that the text-similarity outcome classifier is itself a brittle proxy for substantive compliance change at the editorial-versus-substantive boundary (\S\ref{sec:results-audit}). The within-rulemaking association is statistically inconclusive at this sample size, but the substantive interpretation does not require resolving it: organizations and individuals appear to select into structurally different rulemakings, and the kinds of rules where each population's voice dominates show systematically different revision patterns. This finding is conditional on a permissive reconstruction of commenter type (\S\ref{sec:methods-commenter}).

\paragraph{Contributions.} This paper makes three contributions to the EAAMO research program on equitable access in algorithmic and procedural systems. \textit{Conceptually}, we argue that the relevant unit for auditing administrative responsiveness is not the rule, the docket, or the comment, but the \textit{discrete regulatory obligation}. This unit matches what commenters actually contest and surfaces equity asymmetries that rule-level analysis obscures. \textit{Empirically}, applied to a population-scale EPA corpus, the framework yields three descriptive findings; in both Findings~1 and~3, the data establish an association across rulemakings but not within them. Together these findings reframe the equity question for participatory rulemaking from \textit{``whose comments are heeded?''} to \textit{``which legal obligations do different commenter populations have the resources to identify, interpret, and contest in the first place?''} Formal openness in notice-and-comment rulemaking can mask upstream structural inequity, and obligation-level analysis makes it visible. \textit{Methodologically}, we develop and validate an auditable, LLM-assisted pipeline for obligation extraction and obligation-comment matching at population scale. Blind audits find close agreement with human readers for both components, while the same protocol shows that the text-similarity outcome classifier fails at the editorial-versus-substantive boundary, an error the framework surfaces rather than hides.

Section~\ref{sec:background} situates the paper in the research programs it engages; Sections~\ref{sec:data}--\ref{sec:obligations} describe the corpus, the measurement components, and their validation; Section~\ref{sec:results} presents the results; and Sections~\ref{sec:discussion} and~\ref{sec:limitations} discuss their implications and limitations.

\section{Background}
\label{sec:background}

Our questions sit at the intersection of three established research programs: whether public comments produce substantive change in rule text, and for whom; whether LLM-assisted measurement can reliably code policy-relevant text at population scale; and at what unit of analysis regulatory responsiveness should be measured. Each program is well developed in isolation. Their intersection---population-scale, obligation-level responsiveness measurement with human-audited LLM components---remains underdeveloped.

\paragraph{Public comments and notice-and-comment rulemaking.} Cu\'ellar~\cite{cuellar2005rethinking} develops an early empirical apparatus for measuring whether agency rulemakings substantively consider lay public input, using a five-question sophistication checklist applied to a hand-coded sample. Yackee and Yackee~\cite{yackee2006bias} find that business commenters, but not other commenters, influence the content of final rules. Costa, Desmarais, and Hird~\cite{costa2019public} study whether public comments influence EPA's use of science in National Emission Standards for Hazardous Air Pollutants (NESHAP) rulemakings, finding evidence of differential influence across commenter types. Libgober and Rashin~\cite{libgober2023comments} hand-code comments from high-stakes financial rulemaking and find that most are purely informational, with few threats to involve the courts or Congress. Most recently, Love~\cite{love2026hearing} separates procedural acknowledgment from substantive influence across 854 federal rules from 2017--2023, tracking changes in binding regulatory text, and finds that roughly two-thirds of comment citations come with no responsive change to the rule text. We extend this lineage in two directions: we operate at the obligation level rather than the rule or comment level, and we stratify the analysis by commenter type rather than treating the commenter population as homogeneous.

\paragraph{LLM-as-judge methodology.} A rapidly growing literature evaluates whether large language models can substitute for or assist human content coders in social-science applications. Gilardi, Alizadeh, and Kubli~\cite{gilardi2023chatgpt} show that ChatGPT outperforms crowd workers on standard annotation tasks at substantially lower cost. Heseltine and Clemm von Hohenberg~\cite{heseltine2024large} propose a two-run hybrid (GPT-4 prompted twice, with human adjudication of disagreements) and report 88--95\% accuracy on binary tasks. Pangakis, Wolken, and Fasching~\cite{pangakis2023automated} systematize the case for treating LLMs as annotators when human-coding budgets are limiting, and Ziems et al.~\cite{ziems2024can} survey computational-social-science applications. Recent work emphasizes that LLM annotations require their own validation rather than serving as drop-in replacements for human annotation. Hansen~\cite{hansen2026validating} proposes validating LLM annotations when no reliable human benchmark is available, by testing whether passages an LLM generates from the assigned labels are semantically similar to the original text. Baumann et al.~\cite{baumann2025hacking} show that implementation choices such as model and prompt can push LLM-annotated analyses to incorrect conclusions, call for extra verification of findings near significance thresholds, and find that human annotation protects against false positives.

\paragraph{Computational analysis of regulatory text.} Eidelman and Grom~\cite{eidelman2019argument} introduce a 16-class taxonomy of argument-claim types for public comments at regulations.gov, built by subject-matter experts from about 20,000 comments and applied at scale with weakly supervised labeling. Dong et al.~\cite{dong2026llm} build an LLM-based pipeline for the agency-side counterpart of this task: digesting massive public-comment volumes for agency review. On the rule side, Al-Ubaydli and McLaughlin's RegData~\cite{alubaydli2017regdata} provides the de facto baseline for quantitative regulatory measurement, operationalizing a ``restriction'' as any occurrence in the Code of Federal Regulations of one of five binding-deontic terms (\textit{shall, must, may not, prohibited, required}). Coglianese et al.~\cite{coglianese2021unrules} critique this measure for ignoring ``unrules'' (waivers, exemptions) and for failing to distinguish agency-as-actor from regulated-party-as-actor in deontic sentences. The most direct precedent for our pipeline is Leahey~\cite{leahey2026one}, whose FRTracker corpus operationalizes a ``discrete deontic instruction'' as an (actor, modal, action, object) tuple extracted from binding regulatory text. We inherit Leahey's unit of analysis and his structural extraction approach. Our extension is an LLM verification stage that addresses a limitation Leahey himself acknowledges: purely structural matching cannot distinguish semantic continuity from syntactic continuity when an agency rewrites an obligation without changing its function.

\paragraph{Environmental justice and air-quality regulation.} A substantial public-health and environmental-justice literature documents inequitable exposure to environmental hazards along racial and socioeconomic lines~\cite{mohai2007racial, tessum2019inequity}. EPA rulemakings on air quality, water quality, and toxic substances therefore have stakes that are not uniformly distributed across the population. This distributional context motivates our equity question: whether agency responsiveness to public comments tracks commenter resources rather than the merit or representativeness of the comments themselves.

What remains missing across these programs is a validated measurement framework that links public input to the specific legal obligations agencies actually preserve, revise, or drop. We fill that gap by treating the discrete regulatory obligation as the unit of responsiveness, matching comments to the obligations they address, and auditing whether engagement co-occurs with proposed-to-final obligation outcomes across commenter populations.

\section{Data}
\label{sec:data}

We assemble a population-scale corpus of U.S. EPA public comments submitted to the federal e-rulemaking portal regulations.gov during 2010--2022. The full corpus contains 786,197 comments submitted across 6,145 EPA rulemaking dockets covering air-quality, water-quality, hazardous-waste, pesticide, and toxic-substances regulatory programs (Title 40 of the Code of Federal Regulations [CFR]).

\paragraph{Anchor selection.}\label{sec:data-anchors} For in-depth analysis, we select 36 anchor rulemakings via stratified random sampling combined with extreme-case sampling. Strata cross four year clusters (2010--2012, 2013--2015, 2016--2018, 2019--2022) with comment-volume tertiles (low / mid / high within each year cluster); within each cell, we draw at least one rulemaking at random and supplement with extreme-case selections (highest-volume or otherwise high-attention dockets). The anchors span EPA's major regulatory programs (air quality, water quality, hazardous waste, pesticides, toxic substances), and each anchor rulemaking corresponds to a single regulations.gov docket. Anchor selection favors high-attention rulemakings: the analyzable rate is 32.1\% inside the anchors versus 47.7\% outside, because high-attention dockets disproportionately attract attachment-only and form-letter submissions. Findings on engagement-related patterns therefore apply to this high-attention sample; a 20\% stratified non-anchor sample supplies a partial population baseline (Appendix~\ref{app:schema}). The anchor selection script and the locked anchor table are in the supplementary repository (\href{https://github.com/brucefjn/who-gets-heeded}{\mbox{github.com/brucefjn/who-gets-heeded}}).

\paragraph{Attachment recovery.} Because 56.8\% of EPA comments (446,653 of 786,197) are attachment-only, excluding attachments, as prior work has done~\cite{eidelman2019argument}, would bias the analyzable corpus toward inline submissions and against the many organizational submissions that use PDF attachments. We therefore recover a stratified 1,769-comment attachment sample from within the 36 anchor rulemakings using native extraction (pdfplumber, python-docx) followed by optical character recognition (OCR)~\cite{smith2007overview} with a vision-model fallback for residuals. The pipeline recovers 1,634 of 1,769 attachments (92.4\% within-sample); cross-validation on 141 OCR-recovered comments yields median word-level F1~$= 0.99$ (mean 0.95). Per-stage recovery rates and the OCR cross-validation protocol are in the supplementary repository.

\paragraph{Analytic sample.} Within the 36 anchors, the analytic sample for the responsiveness analysis contains 70,075 analyzable comments: 68,441 free-text comments plus the 1,634 attachment-recovered comments. For each anchor, we pair the proposed and final rules from the Federal Register (72 core documents). Appendix~\ref{app:corpus} describes the broader analytic corpus, the Federal Register pairing, and government documentation of regulations.gov submission practices.

\section{Methods: Measurement components and validation}
\label{sec:methods}

Our findings rest on three computational measurement components, described in \S\ref{sec:obligations}: obligation extraction (LLM-verified), obligation-comment matching (LLM-judged), and outcome-state classification (sentence-transformer cosine similarity). We audited each against blind human judgment (Figure~\ref{fig:pipeline}). The audits found the extraction and matching components to agree closely with human readers (extraction precision $0.95$, $n = 66$; matcher--human Cohen's $\kappa = 0.82$ on addressing and $\approx 0.70$ on stance, $n = 100$), and revealed that the outcome-state classifier fails on the editorial-versus-substantive contrast, which motivates the audit correction of Finding~3 (\S\ref{sec:results-audit}). In all three audits, the two authors coded independently and blind, then adjudicated disagreements. We report two further measurements with their status made explicit: commenter type, a proxy reconstruction that we have not yet audited (below), and a 23-indicator comment-rhetoric schema that we use only for descriptive corpus statistics, not for Findings~1--3 (Appendix~\ref{app:schema}).

\paragraph{Commenter-type reconstruction.}\label{sec:methods-commenter} Equity-stratified analysis (\S\ref{sec:results-equity}) requires labeling each comment as submitted by an \textit{organization} or an \textit{individual}. Because regulations.gov redacts the Organization Name field for personally identifiable information (PII) throughout our bulk-download data, we reconstruct this label from the Title field. The headline analysis uses a permissive title-pattern classifier (it tags titles containing ``submitted by'', ``on behalf of'', or ``written by'' as \textit{individual} and all others as \textit{organizational}), producing a 71.4\% individual / 28.6\% organizational corpus distribution. Because title patterns such as ``on behalf of'' can denote either personal or institutional representation (a community member writing on behalf of a family member vs.\ a corporate representative writing on behalf of an entity), we treat commenter type as a \textit{proxy classification} rather than a validated institutional identity attribute. A stricter ternary classifier with anonymous-detection produces a 14.4\% organizational rate (Appendix~\ref{app:f3-robust}); both classifiers are in the supplementary repository, and we read Finding~3 as conditional on this proxy reconstruction (\S\ref{sec:lim-sampling}).

\section{Methods: Obligation extraction, matching, and outcome classification}
\label{sec:obligations}

To distinguish procedural acknowledgment from substantive change to rule text~\cite{love2026hearing}, we measure responsiveness on the discrete deontic obligation rather than the rule document. The unit of analysis follows Leahey~\cite{leahey2026one}: the meaningful object is the actor-modal-action-object tuple that imposes a binding requirement, not the regulation that contains it.

\paragraph{Extraction.} For each of the 36 anchor rulemakings we extract obligations from both proposed and final Federal Register text. Structural extraction uses spaCy dependency parsing within amendatory blocks, anchored on a closed modal lexicon that distinguishes strong modals (\textit{must, shall, may not, is/are required to, prohibited from}) from permissive modals (\textit{may, is/are permitted to, authorized to}). We extend Leahey's deterministic structural extraction with an LLM verification stage, because agencies routinely preserve a regulatory concept while rewriting its surface form enough to break purely structural matching. The verifier (GPT-5, with closed-enum constraints on every categorical field) also detects compound obligations and emits one structured record per split. A section-level safety-net pass recovers binding obligations the heuristic missed (typically passive constructions and definition-embedded obligations whose modal anchor is implicit) and routes them back through the verifier; on the in-sample baseline (EPA-HQ-OAR-2009-0234, proposed), it recovered 278 verified obligations beyond the heuristic's 569 (a 49\% increase). The full specification, including out-of-scope obligation classes (incorporation-by-reference; pure-definition restatements), is in the supplementary repository.

\paragraph{Obligation-comment matching.} An LLM matcher decides, for each candidate comment-obligation pair, whether the comment substantively \textit{addresses} the obligation and with what \textit{stance} (supporting, opposing, suggesting modification, or none). Candidate pairs require cosine similarity of at least 0.3 between comment and verified obligation and a comment of at least 50 characters. The matcher produces 398,509 candidate pairs from the 70,075-comment analytic sample and labels 20,126 of them (5.05\%) as addressing; at least one commenter addresses 684 distinct obligations (5.6\% of the 12,243 proposed-side obligations). We call obligations addressed by five or more commenters \textit{high-engagement} obligations.

\paragraph{Outcome classification.} We classify each proposed obligation by its fate in the corresponding final rule, using sentence-transformer cosine similarity~\cite{reimers2019sentence} between proposed and final obligation text within the rulemaking: cosine $\geq 0.95 \rightarrow$ SURVIVED-unchanged (SU); $0.85$--$0.95 \rightarrow$ SURVIVED-edited (SE); $0.55$--$0.85 \rightarrow$ MODIFIED (MO); $< 0.55$ or no match $\rightarrow$ DROPPED (DR); final-only obligations with no proposed counterpart $\rightarrow$ NEW. The SURVIVED-edited/MODIFIED boundary operationalizes the procedural-versus-substantive distinction~\cite{love2026hearing} at the obligation level: SURVIVED-edited is an \textit{editorial refinement} of a preserved obligation, and MODIFIED is a \textit{substantive modification}. We call an obligation \textit{revised} if its outcome is SURVIVED-edited, MODIFIED, or DROPPED. Appendix~\ref{app:classifier} reports the classifier specification and its sensitivity to section renumbering.

\paragraph{Validation of extraction and matching.} We report an in-sample precision audit of extraction on a single development rule (40 CFR Part 80 Renewable Fuel Standard [RFS], $n = 66$ verified obligations); held-out validation on additional CFR programs and recall measurement remain outside the present scope. Both authors hand-coded each obligation independently as good, borderline, or bad: inter-annotator agreement was 66/66 exact (Cohen's $\kappa = 1.0$), and precision, counting the 53 good and 10 borderline obligations as correct, is $0.9545$ (95\% Wilson CI~\cite{wilson1927probable} $[0.875, 0.984]$). The three obligations rated \textit{bad} share a single failure mode: the pipeline extracted sentences of the form ``EPA may publish\ldots'' as if they bound regulated parties, when they instead grant discretionary authority to the agency itself, a known limitation of deontic-marker extraction~\cite{coglianese2021unrules}. For the matcher, a blind dual-annotator audit of 100 stratified comment-obligation pairs finds almost-perfect inter-rater agreement (Cohen's $\kappa = 1.000$ on addressing and $0.953$ on stance) and strong agreement between each rater and the matcher (Appendix~\ref{app:matcher}).

\section{Results}
\label{sec:results}

We first report obligation-level responsiveness rates and Findings~1 and~2 (\S\ref{sec:results-rates}), then the commenter-composition analysis behind Finding~3 (\S\ref{sec:results-equity}), and finally the blind audit of the outcome-state classifier on which Finding~3 depends (\S\ref{sec:results-audit}). Table~\ref{tab:findings} summarizes the three findings. Appendix~\ref{app:robust} reports robustness checks, and \S\ref{sec:limitations} states the limitations that constrain each finding.

\begin{table*}[!htbp]
\caption{The three findings. The within-rulemaking estimate for Finding~1 uses docket fixed effects.}
\label{tab:findings}
\Description{A three-row table summarizing the findings. Finding 1: obligations addressed by five or more commenters were revised 69.0 percent of the time versus 51.5 percent for the rest, across 12,243 proposed-side obligations, but the within-rulemaking odds ratio of 1.54 is not significant, p = 0.174. Finding 2: obligations with at least one opposing commenter and those with at least one supporting commenter were revised at similar rates, 62.3 versus 68.3 percent, p = 0.155, across 524 overlapping addressed sets. Finding 3: across rulemakings, the audit-corrected odds ratio is 4.90, p = 0.044, on 115 obligations, compared with a pre-audit odds ratio of 3.07, p = 0.007, on 116.}
\small
\begin{tabular}{lp{0.30\textwidth}p{0.39\textwidth}l}
\toprule
\# & Claim & Main estimate & Scope \\
\midrule
1 & \raggedright Engagement is associated with revision across rulemakings, but not clearly within them & \raggedright 69.0\% vs 51.5\% revised, $p < 0.001$; within-rulemaking OR $= 1.54$, 95\% CI $[0.83, 2.87]$, $p = 0.174$ & 12,243 obligations \\
2 & \raggedright The direction of engagement does not differentiate outcomes & \raggedright 62.3\% (opposing) vs 68.3\% (supporting) revised, $p = 0.155$, in an overlapping-subset comparison & 524 addressed sets \\
3 & \raggedright Across rulemakings, organizational-majority engagement concentrates in editorial refinements & \raggedright Audit-corrected OR $= 4.90$, $p = 0.044$ ($n = 115$); pre-audit OR $= 3.07$, 95\% CI $[1.41, 6.71]$, $p = 0.007$ ($n = 116$) & 116 obligations \\
\bottomrule
\end{tabular}
\end{table*}

\subsection{Obligation-level responsiveness rates}
\label{sec:results-rates}

The extraction pipeline produced 12,730 obligations across the 29 anchor rulemakings with verified extraction output: 12,243 proposed-side obligations and 487 final-only NEW obligations (the other 7 of 36 anchors yield no verified proposed-side obligations; Appendix~\ref{app:classifier}). Across all 12,730 obligations, the outcome distribution is SURVIVED-unchanged 46.4\%, SURVIVED-edited 20.7\%, MODIFIED 27.4\%, DROPPED 1.6\%, and NEW 3.8\%.

\begin{figure}[!htbp]
\centering
\includegraphics[width=0.47\textwidth]{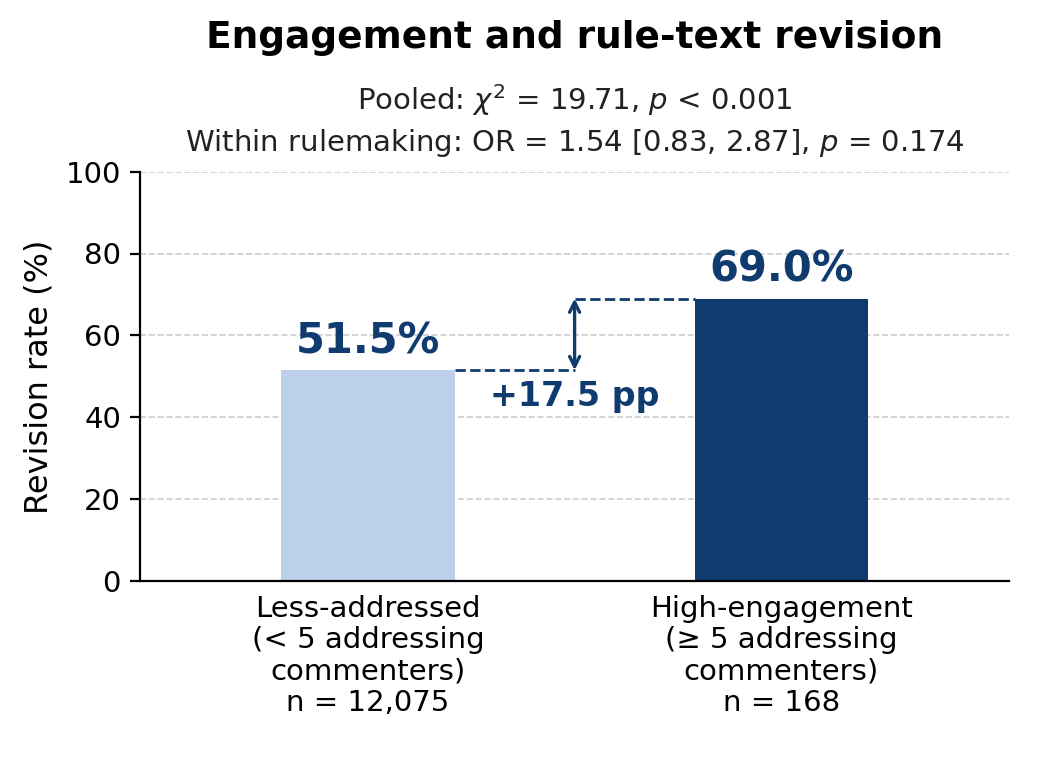}\hfill
\includegraphics[width=0.47\textwidth]{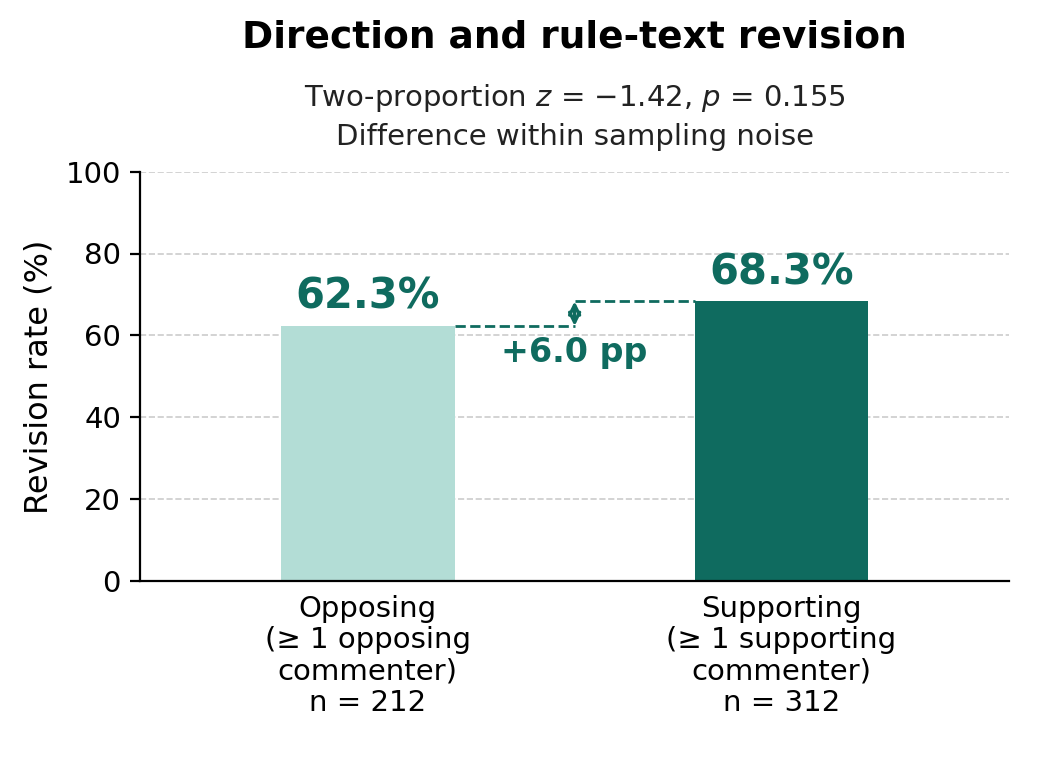}
\caption{Revision rates behind Findings~1 and~2. \textit{Left (Finding~1):} obligations addressed by five or more commenters are revised at 69.0\% versus 51.5\% for less-addressed obligations ($n = 12{,}243$ proposed-side obligations); within the same rulemaking, a docket-fixed-effects logistic regression gives OR $= 1.54$ (95\% CI $[0.83, 2.87]$, $p = 0.174$). \textit{Right (Finding~2):} among obligations with at least one opposing commenter ($n = 212$), 62.3\% are revised; among those with at least one supporting commenter ($n = 312$), 68.3\% are revised. The 6.0-point difference falls within sampling noise (two-proportion $z = -1.42$, $p = 0.155$); given these sample sizes, the null is informative rather than power-deficient.}
\Description{Two bar charts side by side. The left chart compares revision rates by engagement: 51.5 percent revised for obligations with fewer than five addressing commenters and 69.0 percent for obligations with five or more, with a within-rulemaking odds ratio of 1.54 (95 percent confidence interval 0.83 to 2.87). The right chart compares revision rates by direction: 62.3 percent revised for obligations with at least one opposing commenter and 68.3 percent for obligations with at least one supporting commenter, with a two-proportion z of negative 1.42 and p-value 0.155.}
\label{fig:f1-engagement}
\end{figure}

\paragraph{Finding 1: Engagement and revision are associated across rulemakings but not clearly within them.} We test the 12,243 proposed-side obligations: the 487 final-only NEW obligations have no proposed version that could be revised, and none drew an addressing commenter, so we exclude them. The 168 high-engagement obligations were more likely to be revised between proposed and final than less-addressed obligations (69.0\% revised versus 51.5\%; $\chi^2 = 19.71$, $\mathrm{df} = 1$, $p < 0.001$, Cram\'er's $V = 0.040$, $n = 12{,}243$; Figure~\ref{fig:f1-engagement}, left). The small $V$ partly reflects the heavily imbalanced cell distribution. No single docket drives this pooled association: it remains significant in all 29 leave-one-docket-out checks (Appendix~\ref{app:f1}). It also holds under the form-letter, document-pairing, and comment-period checks in Appendix~\ref{app:robust-more} (gap 17.6--21.0 points). To check whether the association holds within rulemakings, we estimate a docket-fixed-effects logistic regression with cluster-robust standard errors (clustered on docket): OR $= 1.54$ (95\% CI $[0.83, 2.87]$, $p = 0.174$; 29 docket clusters). The within-rulemaking association is therefore not statistically distinguishable from no association, and part of the pooled association appears to reflect differences between rulemakings. Most revisions in our corpus still occur on obligations with no addressing commenter.

\paragraph{Finding 2: The direction of engagement does not differentiate outcomes.}\label{sec:results-direction} Among obligations addressed by at least one opposing commenter ($n = 212$), 62.3\% were revised; among those addressed by at least one supporting commenter ($n = 312$), 68.3\% were revised (two-proportion $z = -1.42$, $p = 0.155$; Figure~\ref{fig:f1-engagement}, right). With these sample sizes the analysis would be unlikely to miss a large directional difference; we therefore interpret the null as substantively informative rather than as a power deficit. We caveat the test specification: the two subsets are not strictly disjoint, because an obligation can simultaneously have at least one opposing and at least one supporting commenter, so the two-proportion test treats partially overlapping subsets as independent. A future replication should compare mutually exclusive engagement categories (support-only, opposition-only, mixed, modification-suggesting, no directional stance) in a logistic specification with docket controls; until then, we present the null as substantively informative rather than confirmatory.

\subsection{Commenter composition and outcome type}
\label{sec:results-equity}

\begin{figure}[!htbp]
\centering
\includegraphics[width=3.0in]{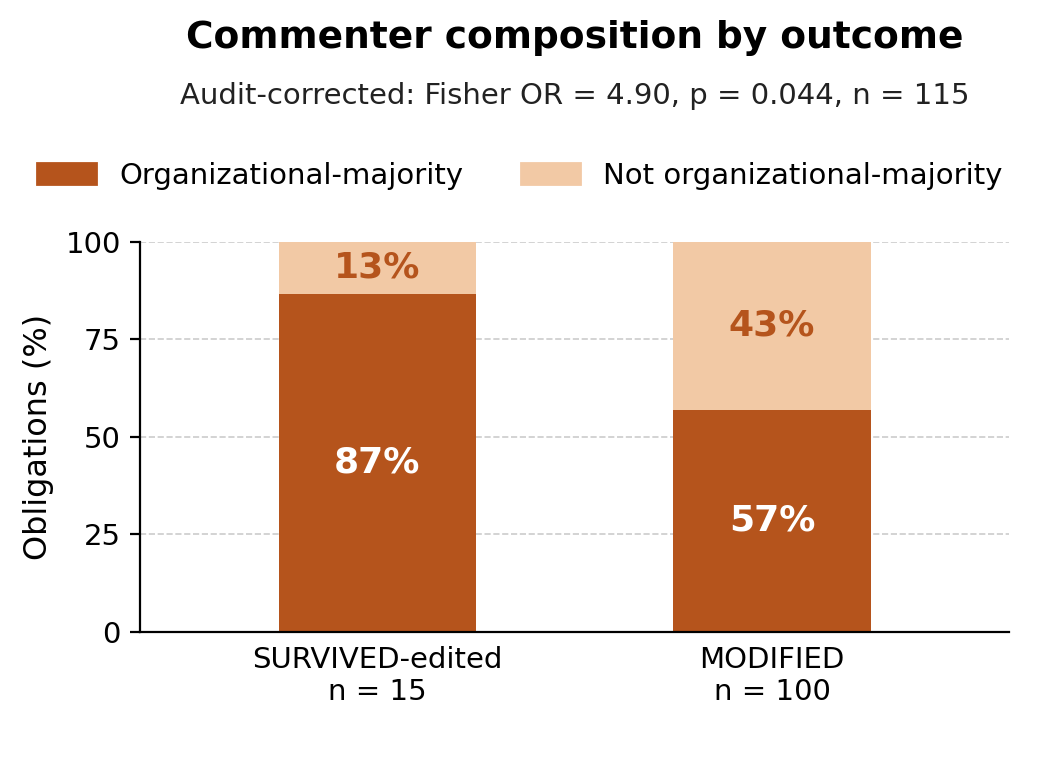}
\caption{Audit-corrected organizational-majority composition by outcome class (Finding~3). On the 115 high-engagement obligations with a SURVIVED-edited or MODIFIED outcome under audit-corrected labels, 87\% of SURVIVED-edited obligations are organizational-majority versus 57\% of MODIFIED obligations. Fisher's exact OR $= 4.90$, $p = 0.044$; pre-audit OR $= 3.07$, 95\% CI $[1.41, 6.71]$, $p = 0.007$ (Table~\ref{tab:f3-contingency}).}
\Description{Two stacked-bar columns comparing organizational-majority composition between SURVIVED-edited and MODIFIED outcomes after audit correction. The SURVIVED-edited column shows 87 percent organizational-majority composition; the MODIFIED column shows 57 percent. A statistical annotation reports a Fisher exact odds ratio of 4.90 with p-value 0.044.}
\label{fig:f3-equity}
\end{figure}

\paragraph{Finding 3: Across rulemakings, organizational-majority engagement concentrates in editorial refinements.} We cross-tabulate commenter composition against outcome type for the 116 high-engagement obligations whose classifier-assigned outcome is SURVIVED-edited or MODIFIED, calling an obligation \textit{organizational-majority} when our classifier labels more than half of its addressing commenters as organizational (\S\ref{sec:methods-commenter}). Under audit-corrected outcome labels (\S\ref{sec:results-audit}), the contingency gives \textbf{Fisher's exact OR $= 4.90$, $p = 0.044$, $n = 115$} (Table~\ref{tab:f3-contingency}, Figure~\ref{fig:f3-equity}; the audit reclassified one obligation as SURVIVED-unchanged). The audit-corrected SURVIVED-edited bucket ($n = 15$) is 87\% organizational-majority, versus 57\% for the MODIFIED bucket ($n = 100$). The audit-corrected result \textit{preserves the direction} of the pre-audit estimate based on classifier-assigned outcomes (OR $= 3.07$, 95\% CI $[1.41, 6.71]$, $p = 0.007$, $n = 116$), with reduced statistical precision (approximate Wald 95\% CI $[1.05, 22.9]$); we read the audit as reducing the concern that the outcome-state classifier alone drives the pattern, not as evidence of a larger effect. The pre-audit association remains significant when each of the 20 dockets with high-engagement obligations is dropped in turn and under a docket-clustered logistic specification (OR $= 3.07$, 95\% CI $[1.08, 8.72]$, $p = 0.035$, 15 docket clusters). Under a section-gated outcome classifier, the direction holds but significance disappears as the SURVIVED-edited bucket shrinks. Under a stricter commenter-type classifier, only 4 of 111 high-engagement obligations are organizational-majority, so the contrast becomes underpowered rather than falsified (Appendix~\ref{app:f3-robust}). Its direction also survives the form-letter, document-pairing, and comment-period checks (pre-audit OR 2.41--4.49; audit-corrected OR 4.18--6.09), although the audit-corrected $p$ reaches 0.07 when only in-period comments count (Appendix~\ref{app:robust-more}). We exclude DROPPED outcomes because none of the 210 DROPPED obligations is high-engagement.

\paragraph{Across rulemakings, not within them.} A Cochran-Mantel-Haenszel (CMH) test~\cite{mantel1959statistical} stratified by docket leaves the within-rulemaking common odds ratio statistically inconclusive (CMH OR $= 1.42$, 95\% CI $[0.51, 4.00]$, $p = 0.507$; 9 strata with 91 observations after dropping degenerate strata; Tarone test of equal odds ratios, $p = 0.139$). We cannot estimate a docket-fixed-effects logistic regression by maximum likelihood because 4 of the 15 contributing dockets show perfect separation; an $L_2$-penalized version yields OR $\approx 2.05$, a biased point estimate without valid inference. We therefore frame Finding~3 as evidence that organizations and individuals select into structurally different rulemakings, and that the kinds of rules where each population's voice dominates correspond to systematically different agency revision behaviors, rather than as evidence of differential agency treatment within a shared rulemaking. The commenter-type proxy remains a separate, unaudited dependency of Finding~3 (\S\ref{sec:lim-sampling}).

\begin{table*}[!htbp]
\caption{Finding~3 contingency before and after audit correction. The audit-corrected SURVIVED-edited bucket shrinks from 58 to 15 obligations and concentrates more sharply on organizational-majority addressing (87\% vs 74\%); the point estimate of the OR rises while the precision of the estimate falls.}
\label{tab:f3-contingency}
\Description{Two rows of counts. Pre-audit: 43 SURVIVED-edited organizational-majority, 15 SURVIVED-edited not organizational-majority, 28 MODIFIED organizational-majority, and 30 MODIFIED not organizational-majority, n = 116, Fisher odds ratio 3.07, p = 0.007. Audit-corrected: 13, 2, 57, and 43, n = 115, odds ratio 4.90, p = 0.044.}
\begin{tabular}{lcccccrr}
\toprule
Contingency & SE-org & SE-not-org & MO-org & MO-not-org & $n$ & Fisher OR & $p$ \\
\midrule
Pre-audit (classifier labels) & 43 & 15 & 28 & 30 & 116 & 3.07 & 0.007 \\
\textbf{Audit-corrected} (adjudicated labels) & \textbf{13} & \textbf{2}  & \textbf{57} & \textbf{43} & \textbf{115} & \textbf{4.90} & \textbf{0.044} \\
\bottomrule
\end{tabular}
\end{table*}

\subsection{Audit of the outcome-state classifier}
\label{sec:results-audit}

Finding~3 turns on the classifier's ability to separate SURVIVED-edited from MODIFIED outcomes. Because the classifier is a cosine-similarity threshold rather than a model of substantive meaning, we cannot assume its validity on this contrast, so we audit it directly.

\paragraph{Sample and procedure.} We audit 150 obligation pairs in two tiers: Tier~1 comprises \textit{all 116} obligations in the Finding~3 contingency, and Tier~2 adds 34 pairs stratified across the remaining outcome classes. Both authors blind-coded independently against a rubric locked before coding began; we excluded cosine similarity, classifier label, and tier from the coding sheet to avoid leakage. The rubric's decision rule for the editorial-versus-substantive contrast asks: \textit{would a regulated party adopt different compliance behavior between proposed and final?} If yes, the outcome is MODIFIED; if no, SURVIVED-edited. The raters disagreed on 6 of 150 pairs (4.0\%), which they adjudicated jointly after blind coding; the rubric, sampler script, and coding sheets are in the supplementary repository.

\paragraph{Inter-rater reliability is almost perfect.} Cohen's $\kappa$ is \textbf{0.898} on the 5-class label (96.0\% exact agreement) and \textbf{0.816} on the SURVIVED-edited versus MODIFIED binary ($n = 134$), both in the almost-perfect range of Landis and Koch~\cite{landis1977measurement} and above the rubric's pre-specified threshold of $\kappa \geq 0.70$ for this contrast. On the Tier~1 subset, $\kappa = 0.816$ (5-class) and $0.803$ (SE/MO). Two independent coders using a locked rubric thus converged without further calibration, evidence that the editorial-versus-substantive distinction is operationalizable (confusion matrix in Appendix~\ref{app:classifier}).

\paragraph{Classifier agreement with human raters is poor on the editorial-versus-substantive contrast.} Against the adjudicated gold standard, the cosine-similarity classifier achieves an SE/MO binary $\kappa$ of only \textbf{0.137} ($n = 129$), ``slight agreement'' under Landis and Koch (5-class $\kappa = 0.324$). The classifier over-calls SURVIVED-edited where human readers identify substantive compliance change: it has high precision but low recall on MODIFIED (precision 0.94, recall 0.52--0.53 against the two raters) and low precision with higher recall on SURVIVED-edited (precision 0.17--0.20, recall 0.61--0.65; Figure~\ref{fig:audit-perclass} in Appendix~\ref{app:classifier}). The cause is interpretable: cosine similarity tracks the preservation of surface text, whereas the SURVIVED-edited/MODIFIED distinction tracks substantive change in compliance behavior. The classifier performs adequately on SURVIVED-unchanged, DROPPED, and NEW (Tier~2 5-class $\kappa = 0.59$); its failure is specific to the editorial-versus-substantive boundary.

\paragraph{What the audit changes.} For Finding~3, the audit shrinks the SURVIVED-edited bucket from 58 to 15 obligations and raises its organizational-majority share from 74\% to 87\% (Table~\ref{tab:f3-contingency}). Findings~1 and~2 are essentially unchanged under audit-corrected labels (Finding~1: $\chi^2 = 18.28$, $p < 0.001$, $V = 0.039$, within-rulemaking OR $= 1.48$, $p = 0.219$; Finding~2: $z = -1.53$, $p = 0.125$), as expected, because both depend on revised-versus-not-revised aggregation rather than on the editorial-versus-substantive distinction. Its main consequence is for Finding~3 and for the framework itself: it exercised the validation hierarchy on a component the findings depend on and produced an interpretable correction. We treat this as evidence that the framework, not the cosine classifier, is the contribution: the classifier is replaceable.

\section{Discussion}
\label{sec:discussion}

Read together, our three findings support one thesis about where equity asymmetries operate in U.S. notice-and-comment rulemaking. Agency revision tracks comment volume only weakly, and not clearly within a rulemaking (Finding~1); it is not differentially responsive to comment direction (Finding~2); and the organizational-versus-individual asymmetry our data support concerns which rulemakings each population engages, while differential treatment within a shared rulemaking remains statistically inconclusive (Finding~3). The asymmetry therefore lies upstream of agency response. Formal openness in notice-and-comment rulemaking can mask upstream structural inequity, and obligation-level analysis makes this pattern visible.

\paragraph{Engagement as an attention signal.} High-engagement obligations are revised more often across the corpus, but not clearly within a given rulemaking. The pooled pattern is consistent with heavily engaged rulemakings also changing more overall; within a rulemaking, our data cannot distinguish a modest effect from none. Public comments may still act partly as an \textit{attention signal}, flagging which obligations matter to a sizable constituency. They do not function as a direct preference-aggregation mechanism that mechanically shifts obligation-level outcomes in the direction of commenter inputs, and the directional null of Finding~2 points the same way. This pattern fits accounts in which forces other than comments drive most revisions, such as review by the Office of Management and Budget (OMB)~\cite{haeder2015influence}, internal technical reconsideration, statutory deadlines, and litigation pressure; it does not fit strong-version responsiveness accounts~\cite{yackee2006bias}. The directional null also admits a second reading: opposing and supporting commenters may address structurally different obligations, so the comparison is partly confounded by selection on what each population engages, the same upstream-selection mechanism that Finding~3 documents for organizational versus individual commenters.

\paragraph{Commenter type and outcome type: evidence of selection across rulemakings.} Organizational-majority commenter composition co-occurs with editorial-refinement outcomes at a substantially higher rate across the high-engagement obligations, under both classifier-assigned and audit-corrected outcome labels. The within-rulemaking estimates are too imprecise to distinguish an undetectably small within-rulemaking effect from a real one this sample cannot characterize (\S\ref{sec:results-equity}). The substantive interpretation does not require resolving this: organizations and individuals appear to select into structurally different rulemakings, and the kinds of rules where each population's voice dominates correspond to systematically different agency revision behaviors. The mechanism our findings point to is therefore \textit{resource asymmetry in engagement allocation}, in which different commenter populations have different capacity to identify, interpret, and contest different legal obligations, rather than differential treatment of organizational versus individual commenters on a shared obligation. We do not claim that our data disprove organizational capture~\cite{yackee2006bias, chen2023administrative} or that individuals are more influential in any causal sense; the data motivate that conversation but cannot settle it without causal identification.

\paragraph{Participation without power.} A growing literature on participation in algorithmic and procedural systems cautions against treating procedural openness as substantive responsiveness. Sloane et al.~\cite{sloane2022participation} and Birhane et al.~\cite{birhane2022power} warn that institutions can invoke participation without redistributing power; Corbett, Denton, and Erete~\cite{corbett2023power} document a tendency in participatory-AI deployments to consult rather than to delegate decision-making weight. Our findings extend this caution to notice-and-comment rulemaking. Every affected party has the same formal right to submit comments, but obligation-level responsiveness requires a more demanding form of access: the capacity to identify which legal obligation is at stake, interpret its consequences, formulate an obligation-specific objection or modification, and do so within the agency's procedural timeline. This is a capacity asymmetry, akin to what Herd and Moynihan call administrative burden~\cite{herd2018administrative}, that the formal right to comment does not address. The equity question for participatory rulemaking is therefore not only who comments, but whose comments can be translated into obligation-specific legal claims that agencies can act upon, and which rulemakings different populations have the resources to engage in the first place.

\paragraph{Implications for transparency infrastructure.} Agencies could report responsiveness at the level of obligations rather than only through preamble summaries of major issues, making agency response legible at the unit at which commenters seek change. Civil-society organizations could use obligation-level tools to help community members identify the deadlines, thresholds, exemptions, monitoring requirements, and reporting duties most relevant to their interests, lowering the capacity barrier our findings document. Audited AI systems can support this kind of accountability infrastructure, but only under a transparent validation hierarchy: extraction, matching, and outcome classification each need their own audit protocols, and researchers should validate the components that findings depend on against human judgment before reporting downstream descriptive or comparative claims~\cite{cen2024transparency}.

\paragraph{What the measurement design makes visible.} Obligation-level measurement reveals patterns that rule-level analysis would obscure: in both Findings~1 and~3, the data establish an association across rulemakings but not within them. We treat the validation hierarchy, which reports each component's audit status, as the methodological contribution most relevant to EAAMO's agenda on algorithmic accountability and equitable institutional participation.

\section{Limitations}
\label{sec:limitations}

The design does not hide its main threats to validity; they are its remaining audit targets. We group them into measurement, sampling, and identification.

\subsection{Measurement limitations}
\label{sec:lim-measurement}

\paragraph{The outcome-state classifier is audited but brittle.} The blind audit (\S\ref{sec:results-audit}) converts Finding~3 into an audit-corrected estimate, but it also shows that the cosine-similarity classifier agrees with humans at only $\kappa = 0.137$ on the SURVIVED-edited/MODIFIED contrast. A natural extension is an LLM-based outcome classifier whose validation reuses this audit's infrastructure.

\paragraph{Recall remains unmeasured.} The matcher sees only candidate pairs above a prefilter threshold (\S\ref{sec:obligations}), so its audit validates classification conditional on retrieval; a below-threshold sample would bound global recall. Recall of obligation extraction is likewise unmeasured: the extraction audit estimates precision on a single development rule, and missed obligations would change the denominator of every rate reported here.

\paragraph{Mass-comment campaigns.} Engagement counts treat each comment as an independent voice, and mass campaigns at EPA could bias the engagement-revision correlation upward~\cite{acus2021mass}. About a third of addressing comments belong to clusters of near-identical form letters, but counting each cluster once does not weaken Finding~1 and preserves the direction of Finding~3 (Appendix~\ref{app:robust-more}).

\paragraph{Few clusters, two raters, and a descriptive schema.} The within-rulemaking and clustered estimates rest on few clusters (29 dockets for Finding~1, 15 for Finding~3), so cluster-robust inference is approximate; a wild-cluster bootstrap~\cite{cameron2008bootstrap} on a larger high-engagement subset is the natural follow-up. All three audits relied on the two authors as coders, who cannot rule out shared blind spots; a larger rater pool would further test robustness. We use the 23-indicator rhetoric schema only descriptively and have not validated it.

\subsection{Sampling and corpus limitations}
\label{sec:lim-sampling}

\paragraph{Anchor selection bias and attachment recovery.} The 36 anchor rulemakings attract more attachment-only and form-letter comments than a random sample (\S\ref{sec:data-anchors}); a 20\% stratified non-anchor baseline shows similar equity-frame prevalence but 19.2 percentage points more explicit opposition inside the anchors (Appendix~\ref{app:schema}). The 1,634 recovered attachments come from a 1,769-comment stratified sample (92.4\% within-sample recovery); pre-2010 dockets and organizational submitters are over-represented in the unrecovered remainder.

\paragraph{Commenter type is a heuristic reconstruction, and Finding~3 is conditional on it.} We reconstruct organizational versus individual identity from Title-field patterns because regulations.gov redacts the Organization Name field throughout the bulk-download data. A conservative ternary classifier (adding anonymous-detection and an ``other'' category) renders the Finding~3 contingency underpowered rather than falsified. We therefore present Finding~3 as \textit{suggestive evidence of upstream engagement asymmetry}. A planned blind validation audit of 200--300 stratified commenter-type labels is the audit required to convert this finding from heuristic-conditional to validated.

\subsection{Identification limitations}
\label{sec:lim-id}

\paragraph{Descriptive, not causal.} All responsiveness estimates are descriptive co-occurrence rates; commenter composition is endogenous to obligation selection, and our patterns are consistent with multiple causal accounts that observational data alone cannot distinguish. Comments are also only one channel of influence: ex parte meetings~\cite{libgober2020meetings}, lobbying before the agency proposes a rule~\cite{yackee2012exparte}, and OMB review~\cite{haeder2015influence} can shape both what agencies propose and what they change, and our design observes only the comment channel. Exploratory splits suggest that the engagement-revision association is concentrated in rulemakings proposed in 2010--2015 and is weakest for rules finalized under the Trump administration (heterogeneity $p \approx 0.04$--$0.05$; Appendix~\ref{app:time}).

\paragraph{Pipeline scope and a small high-engagement subset.} Obligation extraction excludes incorporation-by-reference and some definition-embedded obligations. Seven anchor rulemakings yield no verified proposed-side obligations: six because their Federal Register documents contain no new binding obligations (endangerment findings, methodology meta-rules, recodifications, or binding text without deontic markers) and one because of a recall gap in the heuristic candidate generator (Appendix~\ref{app:classifier}). Commenters address only 5.6\% of proposed-side obligations (684 of 12,243), so the Finding~3 contingency rests on 116 obligations; lowering the addressing threshold or aggregating across anchor cohorts would expand inference power. Headline conclusions are also robust to two specification choices in outcome classification and matcher truncation (Appendix~\ref{app:validation}).

\paragraph{Document pairing.}\label{sec:lim-pairing} The pipeline analyzes each docket's longest proposed and final documents and pools all of the docket's comments. In dockets that host several rulemakings, this produced problematic pairs for four anchors (Appendix~\ref{app:robust-more}), which together supply 9 of the 168 high-engagement obligations. Excluding them, or counting only comments received between the analyzed proposal and final rule, leaves the findings substantively unchanged; re-extracting obligations from verified document pairs is a priority for future work.

\section{Ethics statement}
\label{sec:ethics}

\paragraph{Data source and consent.} The corpus consists exclusively of publicly available federal records: notice-and-comment submissions to regulations.gov and proposed and final rule text from the Federal Register. Comments submitted to a federal rulemaking docket are, by statute, part of the public administrative record. We did not contact commenters, did not augment the public record with any non-public attribute, and did not seek individual consent for inclusion in the corpus, consistent with standard treatment of publicly available regulatory records in administrative-law and computational-policy research.

\paragraph{Personally identifiable information.} The regulations.gov portal redacts the Organization Name field, which is empty throughout the bulk-download data we use. Title-field text frequently contains submitter names; we use Title-field patterns only to construct an aggregate organizational-versus-individual classification and do not surface individual submitter names in the paper, in any released analysis artifact, or in the released audit samples. We release the audit samples without submitter metadata or comment text; comments are identified only by their public regulations.gov IDs. We did not attempt to re-identify any commenter, and we discourage any downstream use of our pipeline that would aim to do so.

\paragraph{LLM-driven measurement and bias risks.} An LLM (GPT-5) produces the obligation-extraction and obligation-comment-matching labels, and its outputs may carry systematic biases inherited from training data. We mitigate this through blind dual-annotator human audits of both measurements and of the outcome-state classifier (\S\ref{sec:obligations}, \S\ref{sec:results-audit}). We did not run cross-model consistency checks against an architecturally distinct verifier or a factorial identity-bias audit of the rhetoric schema in the style of Kim et al.~\cite{kim2026all}; both are natural extensions. The two human raters share an institutional context, so the dual-annotator design alone would not detect systematic shared blind spots.

\paragraph{Misuse and unintended impacts.} In principle, a well-resourced actor could use the co-occurrence patterns we describe between commenter composition and obligation-level outcomes to identify strategies that correlate with favorable outcomes. We mitigate this risk by framing the findings as descriptive co-occurrence rather than as a strategic playbook, by flagging the limits on causal interpretation (\S\ref{sec:limitations}), and by releasing the analytic artifacts (audit samples, reconciliation script, methodology documentation) so that civil-society and academic researchers, not only well-resourced parties, can scrutinize and build on them. We believe the public-interest value of obligation-level transparency outweighs the marginal information benefit our descriptive findings could offer to already well-resourced advocacy operations.

\paragraph{IRB} IRBs typically classify retrospective research on existing public records that contain no identifiable private information as non-human-subjects research; we confirmed this determination with the relevant institutional IRB office prior to data collection.

\paragraph{Reproducibility and contested measurement.} One methodological choice, which we make explicit, materially affects the headline outcome distribution: dropping the same-section gate in outcome classification moves the MODIFIED rate from 61.8\% to 27.4\% (Appendix~\ref{app:classifier}). We preserve the gated classification as a sensitivity artifact and report the substantive findings within the bounded range that the choice implies. Because such choices can substantially shape headline findings, we disclose them in a form that reviewers and downstream users can independently audit.

\bibliographystyle{ACM-Reference-Format}
\bibliography{references}

\appendix

\section{Comment-rhetoric schema and descriptive statistics}
\label{app:schema}

\subsection{Comment-rhetoric feature schema}
\label{sec:methods-features}

We code each comment with 23 theory-grounded yes/no indicators spanning four constructs from the public-comments literature: sophistication of the argument (4 indicators, extending~Cu\'ellar~\cite{cuellar2005rethinking}), explicit stance (2 indicators, support and opposition, allowed to co-occur, following~Eidelman and Grom~\cite{eidelman2019argument}), argument type (12 indicators from~Eidelman and Grom~\cite{eidelman2019argument}: e.g., burdensome, conflicting interests, overreach), and rhetorical frame (5 indicators: technical, legal, justice/equity, economic, lived-experience). Six deterministic features (expert credentials, citation density, regulatory reference counts) enter as covariates. The prompt template and per-indicator operational definitions are in the supplementary repository. Production coding (GPT-5, batch mode) yielded complete labels for 99.987\% of comments (70,066 of 70,075). The schema is used in this paper only for descriptive corpus statistics (Appendix~\ref{sec:results-descriptive}) and the anchor-versus-baseline frame-prevalence comparison (below); the responsiveness Findings~1--3 do not depend on it, and a full per-indicator validation stack is outside the present scope.

\subsection{Descriptive corpus statistics}
\label{sec:results-descriptive}

The 36-anchor analytic sample comprises 70,075 comments, 8.9\% of the 786,197-comment full EPA corpus (2010--2022). Free-text comments account for 68,441 of these (97.7\%); the remaining 1,634 (2.3\%) are attachment-recovered, all from within the 36 anchors. Per-docket comment counts span 71 (EPA-HQ-OAR-2009-0926) to 18,200 (EPA-HQ-OAR-2017-0355) with a heavy right tail, reflecting the anchor-selection methodology's bias toward high-attention rulemakings. The 23-indicator feature schema produced complete labels for 70,066 of 70,075 comments (99.987\%); operational notes on the 9-comment loss (LLM-side schema misses concentrated in rare-indicator fields) and feature-schema compute/cost are in the supplementary repository.

The most-prevalent indicators are: explicit opposition (59.5\%) versus explicit support (25.8\%); the economic / cost-benefit frame (33.1\%) versus justice-equity (18.8\%) and lived-experience (19.4\%); the burdensome and overreach argument types (14.5\%, 10.0\%); and the requested-change sophistication indicator (36.3\%). Per-indicator distributions for all 23 indicators are in the supplementary repository. Explicit opposition outweighs explicit support 2.3-to-1, consistent with the selection-bias literature on commenter incentives~\cite{libgober2020strategic}.

\paragraph{Anchor-sample vs broader-corpus generalization.} A 20\% stratified non-anchor baseline (77,494 comments) shows equity-related frame prevalence (justice-equity, lived-experience) similar to the anchor sample (within 3.2pp), while explicit opposition is 19.2pp higher inside the anchors, quantifying the high-attention-docket selection bias acknowledged in \S\ref{sec:data-anchors}. The 23-indicator instrument is used here descriptively; Findings~1--3 do not depend on it.

\section{Validation details}
\label{app:validation}

\subsection{Obligation-comment matcher audit}
\label{app:matcher}
\label{sec:results-matcher}

The substantive findings rest on two LLM-driven measurements: extraction of binding regulatory obligations from Federal Register text (validated in \S\ref{sec:obligations}: Cohen's $\kappa = 1.0$, precision $= 0.9545$ [Wilson 95\% CI 0.875--0.984], $n = 66$) and the obligation-comment matcher, which classifies which comments substantively address which obligations and with what stance. We validate the matcher with a blind dual-annotator audit on 100 stratified comment-obligation pairs, coding two labels per pair: \textit{addressing} (does the comment substantively engage this specific obligation?) and \textit{stance} (supporting / opposing / suggesting modification / none). Inter-rater agreement is in the almost-perfect range~\cite{landis1977measurement}: Cohen's $\kappa = 1.000$ on addressing and $\kappa = 0.953$ on stance ($n = 100$; $\kappa = 0.897$ restricted to ADDRESSED=TRUE pairs, $n = 49$). Agreement between each human rater and the LLM matcher is strong and symmetric across raters ($\kappa = 0.820$ on addressing for both; $\kappa \approx 0.70$ on stance). The matcher rubric, full inter-rater statistics, and the disagreement list are in the supplementary repository.

\paragraph{Comment truncation.} The matcher initially truncated comment text to 1,400 characters (18.77\% of pairs); a 5,000-character re-run reduces truncation to 1.00\% and shifts the opposition-to-support ratio by 3.8\%. Headline conclusions are robust to this choice.

\subsection{Outcome-state classifier: specification, sensitivity, and audit}
\label{app:classifier}
\label{sec:results-classifier-spec}

Outcome states are assigned by sentence-transformer cosine similarity~\cite{reimers2019sentence} between proposed-rule and final-rule obligation text within each docket: cosine $\geq 0.95 \rightarrow$ SURVIVED-unchanged; $\geq 0.85 \rightarrow$ SURVIVED-edited; $0.55$--$0.85 \rightarrow$ MODIFIED; $< 0.55$ or no match $\rightarrow$ DROPPED; final-only obligations $\rightarrow$ NEW. We do not require a SURVIVED match to occur in the same CFR section: under such a same-section gate, EPA's common practice of renumbering preserved provisions between proposed and final produces an implausible 61.8\% MODIFIED rate (vs 27.4\% under text-only similarity); the two specifications bound the substantive-revision rate at $[27.4\%, 61.8\%]$ and are preserved as a sensitivity-audit artifact in the supplementary repository. The DROPPED-explicit / DROPPED-silent distinction is collapsed because it requires preamble parsing not implemented here. Seven of 36 anchor dockets yield zero verified proposed-side obligations: five are pipeline-correct rejections (endangerment findings, methodology meta-rules, recodifications), one is a legitimate zero (no deontic markers in binding text), and one reflects a recall gap in the heuristic candidate generator, a known limitation; per-docket diagnosis is in the supplementary repository.

\begin{figure*}[!htbp]
\centering
\includegraphics[width=\textwidth]{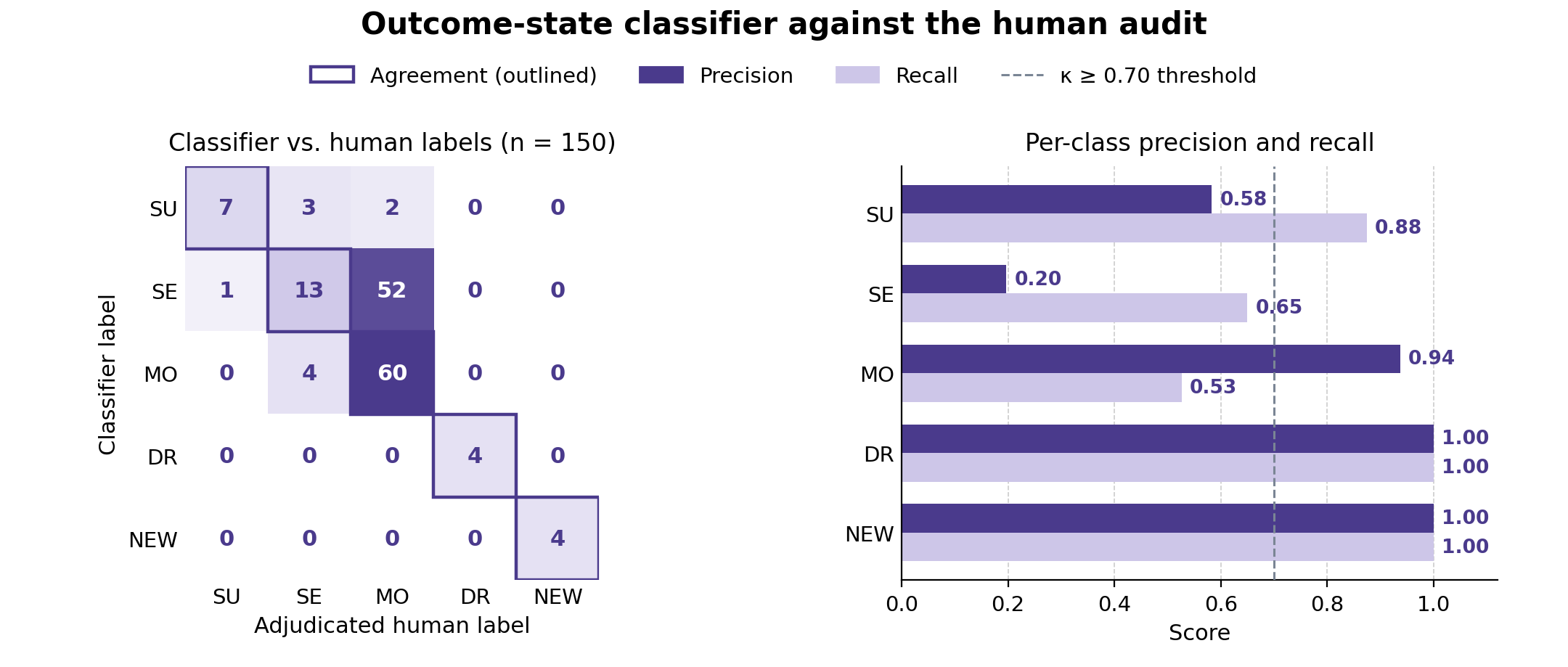}
\caption{Cosine-similarity outcome-state classifier against the human audit ($n = 150$): confusion matrix against the adjudicated labels (left) and per-class precision and recall (right). Performance is strongest on DROPPED and NEW (precision and recall both $1.00$) and acceptable on SURVIVED-unchanged (precision 0.58, recall 0.88); the classifier systematically fails on the SURVIVED-edited / MODIFIED boundary that drives Finding~3, over-calling SURVIVED-edited (low precision 0.20) while missing about half of true MODIFIED obligations (low recall 0.53). The $\kappa \geq 0.70$ line marks the rubric's pre-specified reviewer-defensibility threshold for classifier-vs-human agreement (\S\ref{sec:results-audit}).}
\Description{Two panels. Left: a five-by-five heatmap of classifier predictions (rows) against adjudicated human labels (columns) for 150 obligation pairs; the largest error cell is 52 obligations that the classifier labels SURVIVED-edited and humans label MODIFIED. Right: horizontal bars of per-class precision and recall. SURVIVED-unchanged has precision 0.58 and recall 0.88; SURVIVED-edited has low precision, 0.20, with recall 0.65; MODIFIED has high precision, 0.94, but low recall, 0.53; DROPPED and NEW reach 1.00 on both. A dashed vertical line marks Cohen's kappa of 0.70.}
\label{fig:audit-perclass}
\end{figure*}

\begin{figure*}[!htbp]
\centering
\includegraphics[width=\textwidth]{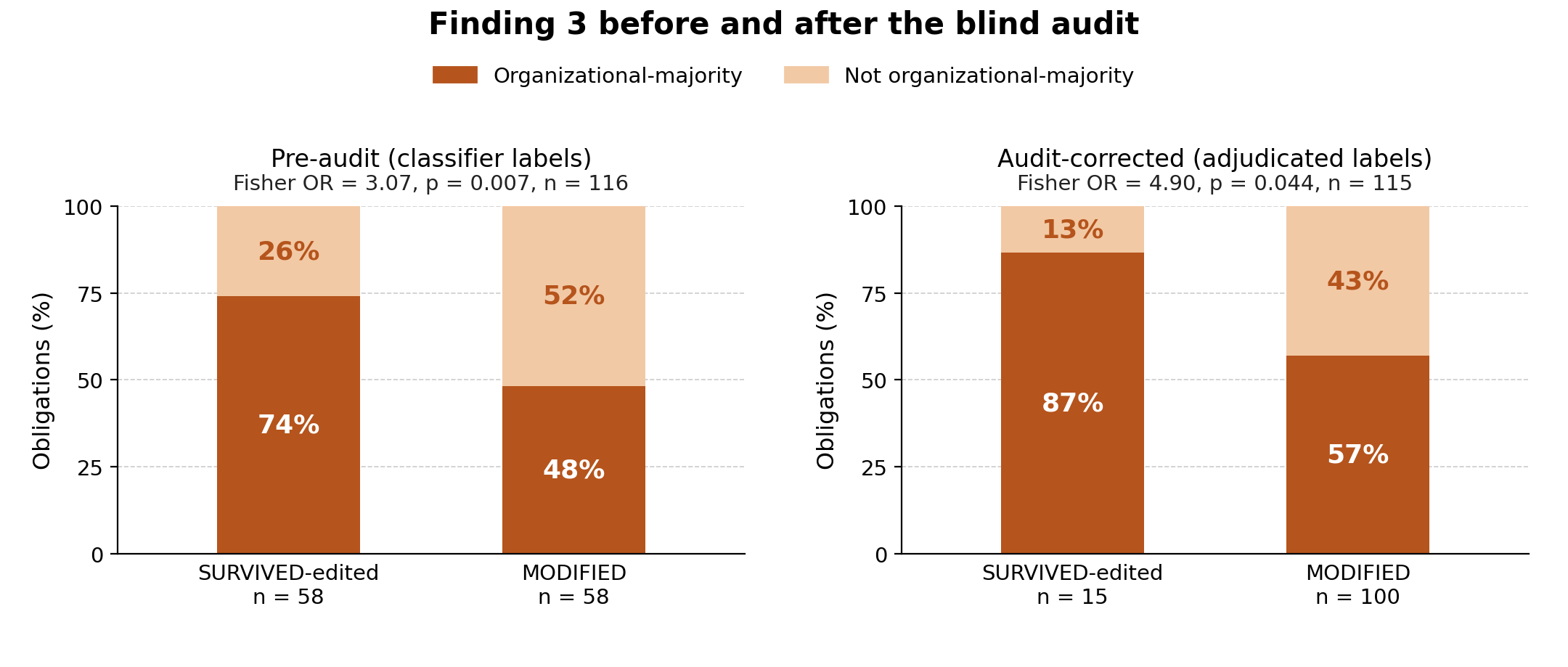}
\caption{Finding~3 organizational-majority composition before and after the blind dual-annotator audit. Pre-audit cells (43, 15, 28, 30) yield Fisher's exact OR $= 3.07$ ($p = 0.007$, $n = 116$); audit-corrected cells (13, 2, 57, 43) yield OR $= 4.90$ ($p = 0.044$, $n = 115$). The audit shrinks the SURVIVED-edited bucket (58 $\to$ 15 obligations) and increases its organizational-majority concentration (74\% $\to$ 87\%), corroborating the directional finding with a larger point estimate while widening the confidence interval.}
\Description{Side-by-side comparison of two stacked bar charts showing organizational-majority composition for SURVIVED-edited and MODIFIED outcome classes, before and after the blind dual-annotator audit. The pre-audit panel shows SURVIVED-edited at 74 percent organizational-majority and MODIFIED at 48 percent. The audit-corrected panel shows SURVIVED-edited at 87 percent and MODIFIED at 57 percent. Statistical annotations show the Fisher exact odds ratio rising from 3.07 to 4.90 with p-values 0.007 and 0.044 respectively.}
\label{fig:f3-audit-comparison}
\end{figure*}

Table~\ref{tab:audit-confusion} reports the inter-rater confusion matrix from the 150-pair audit (\S\ref{sec:results-audit}), Figure~\ref{fig:audit-perclass} reports the classifier's per-class precision and recall against the human audit, and Figure~\ref{fig:f3-audit-comparison} visualizes the Finding~3 contingency before and after audit correction (Table~\ref{tab:f3-contingency}).

\begin{table}[!htbp]
\caption{Inter-rater confusion matrix on the 150-pair audit (rows: rater~1; columns: rater~2). Off-diagonal entries are inter-rater disagreements; the six adjudicated rows split 2 to MODIFIED and 4 to SURVIVED-edited.}
\label{tab:audit-confusion}
\Description{A five-by-five matrix of the two raters' outcome labels on 150 pairs. The raters agree on 8 SURVIVED-unchanged, 16 SURVIVED-edited, 112 MODIFIED, 4 DROPPED, and 4 NEW pairs. The only disagreements are between SURVIVED-edited and MODIFIED: 4 pairs in one direction and 2 in the other.}
\begin{tabular}{lccccc}
\toprule
 & SU & SE & MO & DR & NEW \\
\midrule
SURVIVED-unchanged (SU) & 8  & 0  & 0   & 0 & 0 \\
SURVIVED-edited (SE)    & 0  & 16 & 4   & 0 & 0 \\
MODIFIED (MO)           & 0  & 2  & 112 & 0 & 0 \\
DROPPED (DR)            & 0  & 0  & 0   & 4 & 0 \\
NEW                     & 0  & 0  & 0   & 0 & 4 \\
\bottomrule
\end{tabular}
\end{table}

\section{Robustness checks}
\label{app:robust}

\subsection{Finding 1}
\label{app:f1}

Leave-one-docket-out across the 29 dockets with verified obligations keeps Finding~1's pooled association significant under every condition ($\chi^2$ range $[11.44, 23.77]$, all $p < 0.001$); no single docket drives the pooled association, including EPA-HQ-OPP-2011-0184, which contributes 44 of the 168 high-engagement obligations (full per-docket numbers in the supplementary repository).

\subsection{Finding 3}
\label{app:f3-robust}

Dropping each of the 20 dockets with high-engagement obligations in turn (15 of which contribute to the Finding~3 contingency) keeps the pre-audit Finding~3 significant under every condition (OR range $[2.46, 4.27]$, all $p < 0.05$, CI lower bound always above 1.0); no single docket drives the association (per-docket detail in the supplementary repository). Under the section-gated outcome classification (\S\ref{sec:results-classifier-spec}), the qualitative direction is preserved (OR $= 2.83$, $p = 0.169$, $n = 145$) with significance mechanically lost because the SURVIVED-edited bucket contracts. Under a more conservative ternary commenter-type classifier (anonymous-detection plus an ``other'' category), only 4 of 111 high-engagement obligations are organizational-majority, leaving the contrast underpowered rather than falsified; Finding~3 is therefore conditional on the permissive proxy classifier used in the headline analysis (see \S\ref{sec:lim-sampling}).

\paragraph{DROPPED outcomes.} DROPPED outcomes are not analyzed in the high-engagement cross-tabulation: none of the 210 DROPPED obligations has five or more addressing commenters. DROPPED outcomes occur predominantly on obligations with no addressing commenter (only 3 of the 210, or 1.4\%, have any addressing commenter, versus 5.6\% of proposed-side obligations overall), consistent with drivers outside the comment record (OMB review~\cite{haeder2015influence}, technical reconsideration, statutory deadlines) driving formal removal between proposed and final.

\begin{table*}[!t]
\caption{Robustness of Findings~1 and~3 to form letters, document pairing, and comment periods. Form letters count once as exact-duplicate or near-identical clusters. Flagged dockets are the four listed in Appendix~\ref{app:robust-more}; the comment-period window counts only comments received between the analyzed proposal and final rule. The within-rulemaking odds ratio (OR) uses docket fixed effects; Finding~3 ORs are Fisher's exact, with $p$-values in parentheses and $n$ given as pre-audit/audit-corrected.}
\label{tab:robustness}
\Description{A table of six specifications: the paper's; form letters counted once as exact duplicates; form letters counted once as near-identical clusters; flagged dockets excluded; the comment-period window; and both exclusions together. For Finding 1 it lists high-engagement obligations, the revision-rate gap, and the within-rulemaking odds ratio; for Finding 3 it lists the number of obligations and the pre-audit and audit-corrected odds ratios with p-values. The Finding 1 gap ranges from 17.5 to 21.0 percentage points and the within-rulemaking odds ratio from 1.54 to 1.79, never statistically significant. Finding 3 pre-audit odds ratios range from 2.41 to 4.49 and audit-corrected odds ratios from 4.18 to 6.09, with the audit-corrected p-value between 0.012 and 0.071.}
\small
\begin{tabular}{lcccccc}
\toprule
& \multicolumn{3}{c}{Finding 1} & \multicolumn{3}{c}{Finding 3} \\
\cmidrule(lr){2-4}\cmidrule(lr){5-7}
Specification & High-eng. & Gap (pp) & Within OR [95\% CI] & $n$ & Pre-audit OR & Audit-corrected OR \\
\midrule
Paper & 168 & 17.5 & 1.54 [0.83, 2.87] & 116/115 & 3.07 (0.007) & 4.90 (0.044) \\
Form letters, exact & 164 & 18.0 & 1.57 [0.85, 2.90] & 114/113 & 2.41 (0.036) & 5.75 (0.022) \\
Form letters, near-identical & 149 & 21.0 & 1.79 [0.88, 3.68] & 108/108 & 4.49 ($<$0.001) & 6.09 (0.012) \\
Flagged dockets excluded & 159 & 17.6 & 1.59 [0.84, 3.02] & 109/108 & 2.48 (0.030) & 4.69 (0.044) \\
Comment-period window & 153 & 19.7 & 1.66 [0.92, 2.98] & 109/108 & 3.30 (0.004) & 4.36 (0.070) \\
Excluded and window & 146 & 18.9 & 1.65 [0.91, 2.98] & 102/101 & 2.64 (0.026) & 4.18 (0.071) \\
\bottomrule
\end{tabular}
\end{table*}

\begin{figure*}[!t]
\centering
\includegraphics[width=\textwidth]{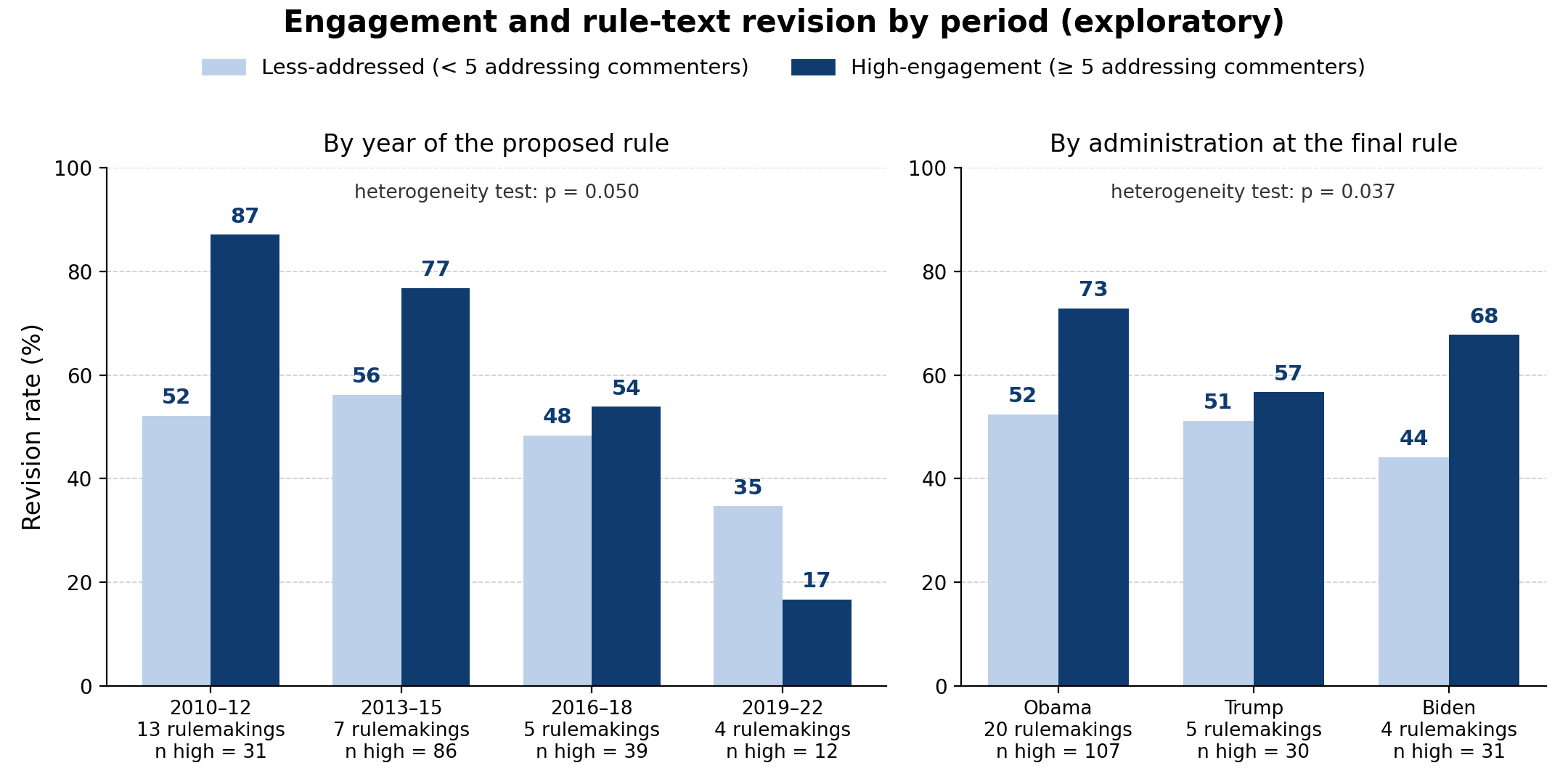}
\caption{Finding~1 by time period (exploratory). Revision rates of high-engagement and less-addressed proposed-side obligations, by the year of the proposed rule (left) and by the administration in office when the analyzed final rule was published (right). Heterogeneity tests are joint Wald tests of high-engagement $\times$ period interactions in a docket-fixed-effects logit with docket-clustered standard errors.}
\Description{Two grouped bar charts of revision rates. Left, by year of the proposed rule: in 2010 to 2012, 52 percent of less-addressed and 87 percent of high-engagement obligations are revised; in 2013 to 2015, 56 and 77 percent; in 2016 to 2018, 48 and 54 percent; in 2019 to 2022, 35 and 17 percent; heterogeneity test p = 0.050. Right, by administration at the final rule: Obama, 52 and 73 percent; Trump, 51 and 57 percent; Biden, 44 and 68 percent; heterogeneity test p = 0.037.}
\label{fig:time-split}
\end{figure*}

\begin{table*}[!t]
\caption{Findings~1--3 by time period (exploratory). Year cluster is the paper's sampling stratum, based on the year of the proposed rule; administration is the one in office when the analyzed final rule was published. Revision rates are for proposed-side obligations, with high-engagement counts in parentheses. Finding~3 cells use audit-corrected outcomes (SE-org/SE-not/MO-org/MO-not).}
\label{tab:time-split}
\Description{A table splitting the findings by the year of the proposed rule and by the administration when the final rule was published. For each period it lists rulemakings and obligations, the revision rates of high-engagement and less-addressed obligations and their gap, the revision rates of obligations with opposing and supporting commenters, and the audit-corrected Finding 3 cells. The Finding 1 gap is 35.0, 20.6, 5.5, and minus 18.0 percentage points for 2010 to 2012, 2013 to 2015, 2016 to 2018, and 2019 to 2022, and 20.5, 5.6, and 23.7 points for the Obama, Trump, and Biden administrations.}
\small
\begin{tabular}{lcccccc}
\toprule
& & \multicolumn{3}{c}{Finding 1: revised (\%)} & Finding 2: revised (\%) & Finding 3 \\
\cmidrule(lr){3-5}
Period & Rulemakings / obligations & High-eng. & Less-addressed & Gap (pp) & Opposing / supporting & Cells \\
\midrule
\multicolumn{7}{l}{\textit{Year cluster of the proposed rule}} \\
2010--2012 & 13 / 7,075 & 87.1 (31) & 52.1 & 35.0 & 78.0 / 65.0 & 3/0/7/17 \\
2013--2015 & 7 / 3,229 & 76.7 (86) & 56.2 & 20.6 & 65.9 / 81.0 & 7/2/35/22 \\
2016--2018 & 5 / 989 & 53.8 (39) & 48.3 & 5.5 & 50.0 / 53.8 & 3/0/15/2 \\
2019--2022 & 4 / 950 & 16.7 (12) & 34.6 & $-$18.0 & 27.3 / 35.0 & 0/0/0/2 \\
\midrule
\multicolumn{7}{l}{\textit{Administration when the final rule was published}} \\
Obama & 20 / 10,095 & 72.9 (107) & 52.4 & 20.5 & 64.1 / 68.2 & 13/0/37/28 \\
Trump & 5 / 942 & 56.7 (30) & 51.1 & 5.6 & 50.9 / 54.2 & 0/0/14/2 \\
Biden & 4 / 1,206 & 67.7 (31) & 44.1 & 23.7 & 82.4 / 83.6 & 0/2/6/13 \\
\bottomrule
\end{tabular}
\end{table*}

\subsection{Form letters, document pairing, and comment periods}
\label{app:robust-more}

Engagement counts treat every comment as an independent voice. To test whether form-letter campaigns drive the results, we cluster the 6,513 comments that address at least one obligation within each docket, either as exact duplicates after normalizing the text or as near-identical texts (Jaccard similarity $\geq 0.7$ on three-word shingles, found with MinHash and verified exactly), and count each cluster once per obligation; a cluster takes the majority commenter type of its members. About a third of addressing comments (34.3\%) belong to multi-member near-identical clusters; the largest is a 608-comment campaign in EPA-R08-OAR-2015-0463. We also check the document pairing (\S\ref{sec:lim-pairing}). For four anchors, the rule of analyzing each docket's longest proposed and final documents produced problematic pairs: a 2013 proposal matched with a 2024 final rule (EPA-HQ-OW-2009-0819), a final rule that predates its proposal (EPA-HQ-OAR-2002-0058), a proposal published after the comment corpus ends (EPA-HQ-OAR-2015-0072), and identical documents assigned to two anchors (EPA-HQ-OAR-2009-0234 and EPA-HQ-OAR-2011-0044, of which we keep the first). Together they supply 9 of the 168 high-engagement obligations and 7 of the 116 in the Finding~3 contingency. We first exclude these four dockets and then count only comments received between the analyzed proposal and final rule, which keeps 18,235 of the 20,126 addressing comment--obligation pairs. Table~\ref{tab:robustness} reports both findings under each specification. The Finding~1 gap and within-rulemaking odds ratio are unchanged or slightly larger throughout, and the within-rulemaking association remains statistically inconclusive. Finding~3 keeps its direction throughout; its magnitude varies with the specification, and its audit-corrected $p$-value is about 0.07 when only comments from the analyzed comment period count.

\subsection{Time periods}
\label{app:time}

Table~\ref{tab:time-split} and Figure~\ref{fig:time-split} split the three findings by the paper's sampling strata, which are based on the year of the proposed rule, and by the administration in office when the analyzed final rule was published. The Finding~1 gap is largest in rulemakings proposed in 2010--2015 and is small or reversed in those proposed later; by administration, it is smallest for rules finalized under the Trump administration. A docket-fixed-effects logit with high-engagement $\times$ period interactions and docket-clustered standard errors gives within-rulemaking odds ratios of 6.54, 1.95, 0.78, and 0.19 across the four year clusters (joint Wald test $\chi^2(3) = 7.80$, $p = 0.050$) and of 2.22, 0.37, and 1.55 for the Obama, Trump, and Biden administrations ($\chi^2(2) = 6.59$, $p = 0.037$). The Finding~2 null holds in every period except 2013--2015 ($p = 0.009$). Under audit-corrected labels, all 13 organizational-majority editorial refinements come from rulemakings finalized under the Obama administration. Later periods contain only four or five rulemakings each, and the Biden-era group includes EPA-HQ-OW-2009-0819 (\S\ref{sec:lim-pairing}), so these splits are exploratory.

\section{Corpus construction}
\label{app:corpus}

The final analytic corpus is 340,910 comments, comprising 339,276 free-text comments (regulations.gov inline text from across the 6,145-docket corpus) and 1,634 attachment-recovered comments (all from within the 36 anchor rulemakings). Within the 36 anchors specifically (the analytic sample for our responsiveness analysis), we have 70,075 analyzable comments (68,441 free-text plus the 1,634 attachment-recovered). The 23-indicator feature schema (Appendix~\ref{app:schema}) codes all 70,075 analytic-sample comments. The obligation-comment matcher (\S\ref{sec:obligations}) applies an additional eligibility floor (cosine similarity to a verified obligation at or above 0.3 and a 50-character minimum comment length), yielding a smaller pair-eligible subset reported in \S\ref{sec:results}. Feature-schema distributions and matcher responsiveness rates are therefore computed over different denominators.

For each of the 36 anchor rulemakings, we retrieve both the proposed rule and final rule from the Federal Register API. Regulation Identifier Number (RIN) pairing is verified via preamble cross-reference checks. All 36 anchors have a matched proposed--final Federal Register document pair (72 core documents); we also retrieved 334 candidate and related Federal Register documents (interim rules, technical corrections, withdrawals) during the pairing process for traceability.

Government-side documentation of regulations.gov submission practices and PII redaction conventions that constrain the commenter-type reconstruction (\S\ref{sec:methods-commenter}) is in the relevant GAO reports~\cite{gao2019, gao2020, gao2021}.

\end{document}